\documentclass[runningheads]{llncs}
\usepackage{graphicx}
\usepackage{subfigure}
\usepackage{amsmath}
\usepackage{amssymb}
\usepackage{amsfonts}
\usepackage{multirow}
\usepackage[compatibility=false]{caption}% http://ctan.org/pkg/caption
\usepackage{pifont}
\usepackage[noadjust]{cite}
\DeclareMathOperator*{\argmax}{arg\,max}
\DeclareMathOperator*{\argmin}{arg\,min}
\usepackage{caption}

\usepackage{booktabs}

\begin{document}

\title{Signed Laplacian Deep Learning with Adversarial Augmentation for Improved Mammography Diagnosis}
\titlerunning{Signed Laplacian Deep Learning with Adversarial Augmentation}
\author{Heyi Li\inst{1}\thanks{These authors contribute equally to this work.} % index{Li, Heyi} 
	\and
	Dongdong Chen\inst{1}\textsuperscript{*} % index{Chen, Dongdong} 
	\and 
	William H. Nailon\inst{2} % index{Nailon, William}
	\and \\ Mike E. Davies\inst{1} % index{Davies, Mike}
	\and David I. Laurenson\inst{1} % index{Laurenson, David}
	}
\authorrunning{H. Li et al.}
\institute{Institute for Digital Communications, University of Edinburgh, Edinburgh, UK \\
	\email{\{Heyi.Li,d.chen, dave.laurenson, mike.davies\}@ed.ac.uk} \and
	Oncology Physics Department, Edinburgh Cancer Centre, Western General
	Hospital, Edinburgh
	\email{bill.nailon@luht.scot.nhs.uk}}
\maketitle             

\begin{abstract}
Computer-aided breast cancer diagnosis in mammography is limited by inadequate data and the similarity between benign and cancerous masses. To address this, we propose a signed graph regularized deep neural network with adversarial augmentation, named \textsc{DiagNet}. Firstly, we use adversarial learning to  generate positive and negative mass-contained mammograms for each mass class.
After that, a signed similarity graph is built upon the expanded data to further highlight the discrimination. 
Finally, a deep convolutional neural network is trained by jointly optimizing the signed graph regularization and classification loss. Experiments show that the \textsc{DiagNet} framework outperforms the state-of-the-art in breast mass diagnosis in mammography.

\end{abstract}

\section{Introduction}
\label{Introduction}
Breast cancer is one of the most frequently diagnosed mortal diseases for women all over the world \cite{breaststa2014}. Mammography is extensively applied and computer-aided diagnosis systems (CADs) are often employed as a second reader. Leveraging the recent success of deep neural networks on representation learning, deep learning based CADs \cite{dhungel2016automated, zhu2017deep, shams2018deep,wu2018conditional,li2019deep, li2018improved} outperform traditional methods, which rely heavily on handcrafted features. However, two major challenges in mammographic CADs remain (1): limited access to well annotated data \cite{dhungel2016automated} and (2): the similarity between benign and cancerous masses. To alleviate the impact of inadequate data, \cite{dhungel2016automated, zhu2017deep, li2019deep, li2018improved} applied classical geometric transformations for data augmentation (e.g. flips, rotations, random crops etc), and more recently, \cite{shams2018deep,wu2018conditional} generated synthetic images on the manifold of real mammograms using adversarial learning \cite{goodfellow2014generative}, which enjoys a powerful ability to learn the unknown underlying distribution. 
Unfortunately, the following questions remain unanswered: What kind of data augmentation is most helpful for CADs in mammography? How can we alleviate the impact of the similarity between data, i.e., how can we maximize the margin between manifolds with a small difference?

In this paper, we propose a new deep learning framework that improves mammography diagnosis as follows. Firstly, we propose an adversarial data augmentation strategy, in which both positive and negative samples of specific classes are generated in an unsupervised manner, in order to make more distinct boundaries between different classes. 
After that, we build a signed graph Laplacian over the augmented data to quantitatively capture the geometric structure of data. Finally, we train a deep neural network by jointly optimizing the graph regularization and classification loss, by which the intra-class difference is minimized, and more importantly, the inter-class manifold margin is maximized in the deep representation space. Extensive experiments show that the proposed \textsc{DiagNet} outperforms the state-of-the-art of breast masses diagnosis in mammography.

\section{Preliminary}
\subsection{Adversarial learning}
Adversarial learning is a technique that attempts to fool models through malicious input \cite{kurakin2017adversarial} and has achieved impressive results in representation learning. 
%The key idea of its success is to force the output of the generator to be indistinguishable from the real input, which is derived from a Generative Adversarial Network (GAN) \cite{goodfellow2014generative}. 
The key idea of the success of a Generative Adversarial Network (GAN) is to force the output of the generator to be indistinguishable from the real input \cite{goodfellow2014generative}.
Adversarial training is particularly powerful for image generation, and for learning unknown and complicated distributions from the training data. In this paper, we propose to use adversarial learning to generate off-distribution instances along with on-distribution instances in order to enlarge the medical image data.

\subsection{Manifold Learning}
In real applications, data typically reside on a low-dimensional manifold embedded into a high-dimensional ambient space \cite{seung2000manifold}. Manifold learning is extensively explored  because of its effectiveness for preserving the topological locality, which relies on the assumption that neighbors tend to have the same labels \cite{chen2017unsupervised}. In this paper, we aim to incorporate graph embedding into a deep neural network as a regularizer in the latent space. In addition, local data manifold structure preservation within the hidden representations in deep neural networks offers the possibility of improving the performance of the classifier \cite{chen2018graph}.

\section{Proposed Method}
\begin{figure}[t]
\begin{center}
    \subfigure[]{\includegraphics[width=0.631\linewidth]{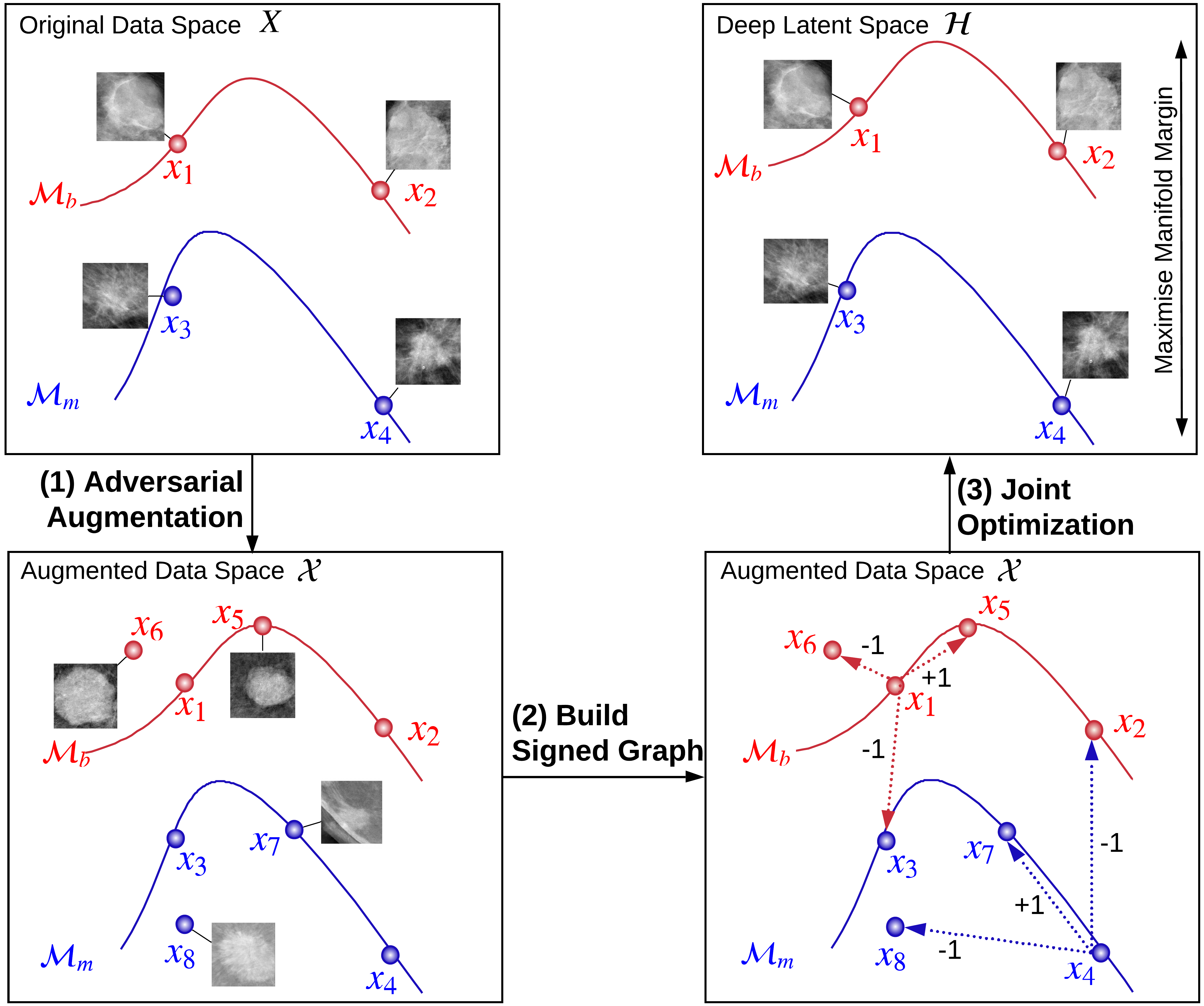}
    \label{fig:motivations} }
    \hspace{2ex}
    \subfigure[]{\includegraphics[width=0.1262\linewidth]{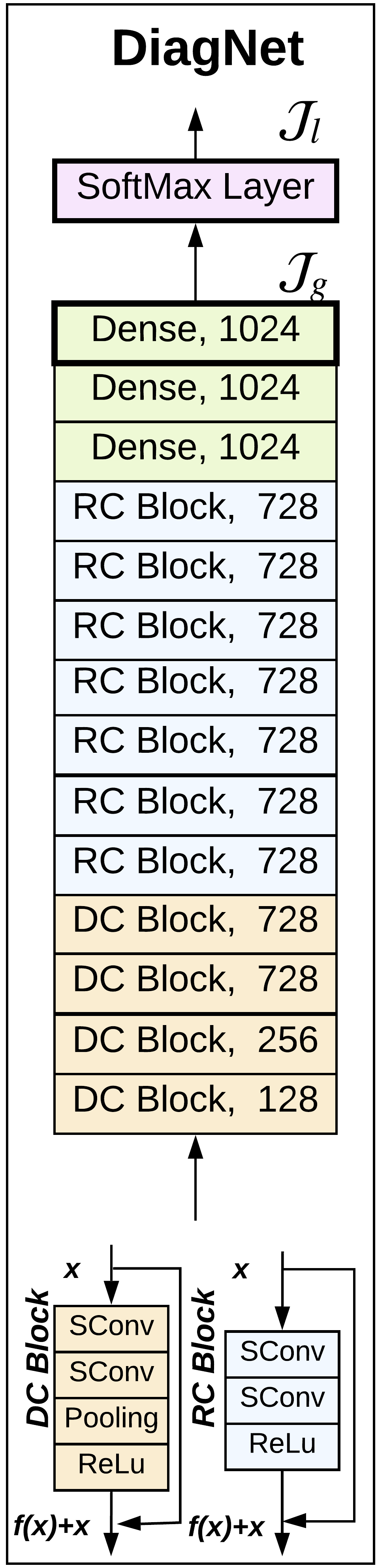}
    \label{fig:network} }
\end{center}
\vskip -0.2in
\caption{The proposed \textsc{DiagNet} for Breast Mass Diagnosis. (a) the framework of the proposed algorithm, which consists of three steps. $\{x_1, x_2\}$ and $\{x_3, x_4\}$ are samples on benign manifold $\mathcal{M}_b$ and malignant manifold $\mathcal{M}_m$ respectively. In the first step, i.e. adversarial data augmentation, positive neighbors $\{x_5, x_7\}$ and negative neighbors $\{x_6, x_8\}$ are generated with (\ref{eq_positiveneighbors}) and (\ref{eq_negativeneighbors}) respectively. Then a signed graph is built upon both original and augmented samples as (\ref{eq_adjacency}). Finally, a joint loss (\ref{eqs:loss}) is optimized in the deep latent space, maximizing data manifold margin. (b) The utilized deep network architecture. ``DC block" represents a down-sampling convolutional block, ``RC block" is a residual convolutional block, and ``SConv" is separable convolutions.}
\label{fig:flowchart}
%\vskip -0.18in
\end{figure}

In this section, we formally introduce the details of \textsc{DiagNet}, which is composed of three steps as shown in Fig.\ref{fig:flowchart}: (1) adversarial augmentation, (2) a signed graph Laplacian built upon the augmented data and (3) joint optimization of the classifier loss and signed graph regularizer. 
We first define the notation applied throughout the paper. Let $\{X, Y\}=\{\boldsymbol{x}_i, y_i\}_{i=1}^{n}$ be the $n$ mammograms with corresponding labels, where $\boldsymbol{x}_i \in \mathbb{R}^{H\times W}$ is an image sample and $y_i \in \{y_c\}_{c=1}^C$ is the class label. Let $\{X_c, y_c\}$ denote the $c$-th class data.

\subsection{Adversarial Augmentation}
\label{sec:augmentation}
As also mentioned in section \ref{Introduction}, inadequate data and the similarity between benign and cancerous masses \cite{dhungel2016automated} are two main reasons causing high false positives in mammographic CADs. Recently, \cite{antoniou2017data, wu2018conditional, shams2018deep} employed GANs to create new instances. Even though they generated on-distribution samples that are not separable by discriminators, they ignored the importance of distinguishable but similar instances, which tend to improve the discriminative ability. To overcome this shortcoming, as shown in Fig.\ref{fig:motivations}, we propose to use adversarial learning to generate more instances  of both \textit{positive neighbors} (i.e. instances on the manifold, e.g. $x_5$ and $x_7$) and \textit{negative neighbors} (i.e. instances off the manifold, e.g. $x_6$ and $x_8$). Here, there are defined two manifolds: $\mathcal{M}_b$ for benign images and $\mathcal{M}_m$ for malignant images.

In particular, inspired by \cite{yu2017open}, we generate neighboring instances one by one for a certain data class $\{X_c, y_c\}$, $c = 1, 2, \cdots, C$, where $C=2$ in this paper. Specifically, both positive and negative neighbors are generated based on the noise corrupted seed points (a number of randomly selected samples in $X_c$) and they are both close to the original data points. In particular, the positive neighbors ${X_c^+}$ are the generated samples that cannot be separated from $X_c$ by a discriminator, while the negative neighbors ${X_c^-}$ are the ones that can be separated. Finally, the expanded dataset for class $c$ is of the form  $\mathcal{X}_c=\{X_c \cup X_c^+ \cup X_c^-\}$, and the whole dataset is $\mathcal{X}=\bigcup_c\mathcal{X}_c$.

Let $\boldsymbol{x}$ be a desired new sample and $P(\boldsymbol{x}; {X}_{c}, {X}_c^+)$ be the probability that $\boldsymbol{x}$ is classified as class $c$ by a discriminator trained on $\{X_c, X_c^+\}$. Similarly $P(\boldsymbol{x}; {X}_{c}, {X}_c^-)$ corresponds to a discriminator trained on $\{X_c, X_c^-\}$.
Note that $X_c^+$ and $X_c^-$ are initialized as empty. In this paper, we trained two SVM classifiers as the discriminators and the corresponding output probability is obtained with logistic sigmoid of the output signed distance. Accordingly, a set of neighboring instances $\{\boldsymbol{x}_t\}_{t=1}^T$ of $X_c$ are iteratively generated. In each iteration $t$, the discriminator is learned and the weights are updated. After $T$ iterations of training, we select one desired positive neighbor $\boldsymbol{x}$:
\begin{equation}
\argmax_{\boldsymbol{x}} P\big(\boldsymbol{x}; X_c, X_c^+\cup \{\boldsymbol{x}_t\}_{t=1}^T\big) - \gamma \max \{0, r_1 - \min_{\boldsymbol{x}_i \in X_c^+} d(\boldsymbol{x}, \boldsymbol{x}_i)\}, 
\label{eq_positiveneighbors}
\end{equation}
where $d(\cdot)$ is a distance measure, $\gamma$ weights the distance regularization, forcing generated points to be different with a minimum distance $r_1$. Similarly, we select one desired negative neighbor $\boldsymbol{x}$, with an added distance restriction to force new points to be scattered close to $X_c$:
\begin{equation}
\begin{aligned}
\argmin_{\boldsymbol{x}} P\big(\boldsymbol{x}; X_c, X_c^-\cup \{\boldsymbol{x}_t\}_{t=1}^T\big) &+ \gamma \max \{0, r_2 - \min_{\boldsymbol{x}_j \in X_c^-} d(\boldsymbol{x}, \boldsymbol{x}_j)\} 
\\
&+
\gamma \max \{0, \min_{\boldsymbol{x}_i \in X_c} d(\boldsymbol{x}, \boldsymbol{x}_i) - r_3\}, 
\label{eq_negativeneighbors}
\end{aligned}
\end{equation}
where the distance regularization forces generated points to acquire a minimum distance $r_2$ and maximum distance $r_3$.

\subsection{Signed graph Laplacian regularizer}
Graph embedding trained with distributional context can boost performance in various pattern recognition tasks. In this paper, we aim to incorporate the signed graph Laplacian regularizer \cite{chen2018learning} to learn a discriminative datum representation $\mathcal{H}(\mathcal{X})$ by a deep neural network, where discriminative here means that the intra-class data manifold structure is preserved in the latent space and the inter-manifold (slightly different) margins are maximized.

Using the supervision of the adversarial augmentation in section \ref{sec:augmentation}, we build a signed graph upon the expanded data $\mathcal{X}$.
Given $\mathcal{X}_c = \{X_c, X_c^+, X_c^-\}$ for class $c$,  and all other classes data
$\mathcal{X}_{-c} = \bigcup_{t=1,\cdots,C\\; t\neq c}\{X_t, X_t^+, X_t^-\}$, 
% $\mathcal{X}_{-c} = \{X_{-c}, X_{-c}^+, X_{-c}^-\}$, 
for $\forall \boldsymbol{{x}}_i \in \mathcal{X}_c$, the corresponding elements in the signed graph is built as follows:

\begin{equation}
\label{eq_adjacency}
\phi_{ij}=\begin{cases}
+1,  & \boldsymbol{{x}}_j \in \{X_c \cup X_c^+\}_i^{n^+}, \\
-1,  & \boldsymbol{{x}}_j \in \{X_{-c} \cup \mathcal{X}_c^-\}_i^{n^-}, 
\end{cases}
\end{equation} 
where the $\{\cdot\}_i^{n^+}$($\{\cdot\}_i^{n^-}$) denotes the corresponding $n^+$ ($n^-$) nearest neighborhood of $x_i$ to approximate the locality of the manifold.

Then, we compute the structure preservation in the deep representation space (directly behind the softmax layer as shown in Fig.\ref{fig:network}) $\mathcal{H}=\{{h}(\boldsymbol{x}_i)\}_{i=1}^N$, where $N = |\mathcal{X}|$. The signed graph Laplacian regularizer is defined as following:
\begin{equation}\label{eqs:E_graphMatrix}
\mathcal{J}_g(\mathcal{X}, \Phi) = 
\sum\limits_{i,j} \begin{cases}
\phi_{ij} \cdot dist({h}(\boldsymbol{x}_i), {h}(\boldsymbol{x}_j)),   &\text{if } \phi_{ij} > 0\\
\max \big(0, m + \phi_{ij} \cdot dist({h}(\boldsymbol{x}_i), {h}(\boldsymbol{x}_j)) \big),  &\text{if } \phi_{ij} < 0,
\end{cases}
\end{equation}
where $dist(\cdot)$ is a distance metric for the dissimilarity between ${h}(\boldsymbol{x}_i)$ and ${h}(\boldsymbol{x}_j)$. It encourages similar examples to be close, and those that are dissimilar to have a distance of at least m each other, where $m>0$ is a margin.

Note that instead of calculating the manifold embedding by solving an eigenvalue decomposition, we learn the embedding $\mathcal{H}$ by a deep neural network. Specifically, inspired by the depth-wise separable convolutions \cite{chollet2017xception} that are extensively employed to learn mappings with a series of factoring filters, we build stacks of depth-wise separable convolutions with similar topological architecture to that in \cite{chollet2017xception} to learn such deep representations (Fig.\ref{fig:network}).

Therefore, by minimizing (\ref{eqs:E_graphMatrix}), it is expected that if two connected nodes $\boldsymbol{{x}}_i$ and $\boldsymbol{{x}}_j$ are from the same class (i.e. $\phi_{ij}$ is positive), ${h}(\boldsymbol{x}_i)$ and ${h}(\boldsymbol{x}_j)$ are also close to each other, and vice versa. Benefiting from such learned discriminativity, we train a simple softmax classifier to predict the class label, i.e.,
\begin{equation}\label{eqs:label_crossentropy}
\mathcal{J}_l = -\frac{1}{N}\sum_{i=1}^N\sum_{c=1}^C
{\delta_c(y_i)\log P\big(y_i\mid \boldsymbol{{x}}_i;\boldsymbol{\theta} \big)},
\end{equation}
where $\delta_c(y_i)=1$ when $y_i=c$, and $0$ otherwise; $\boldsymbol{\theta}$ is the parameter set of the neural network. 

Finally, by incorporating the signed Laplacian regularizer (\ref{eqs:E_graphMatrix}) and  the classification loss  (\ref{eqs:label_crossentropy}), the total objective of \textsc{DiagNet} is accordingly defined as:
\begin{equation}\label{eqs:loss}
\mathcal{J} =   \mathcal{J}_l + \lambda\mathcal{J}_g,
\end{equation}
where $\lambda \geq 0$ is the regularization trade-off parameter which controls the smoothness of hidden representations.

\section{Experiments}
\label{sec:exp}
\subsection{Datasets and ROIs selection}
\label{ssec:data}
The \textsc{DiagNet} is evaluated on the most frequently used full-field digital mammographic dataset, INbreast \cite{moreira2012inbreast}. 107 mass contained mammograms are divided into a training and a test set containing 80\% and 20\% of the images respectively.  As for ROIs selection, rectangular mass-contained boxes are selected with proportional padding  ($1.6$ times) upon original ROI bounding boxes. The selected ROIs are augmented with flips and further adversarially augmented by 40\% more (20\% positive neighbors and 20\% negative neighbors).

\subsection{Implementation Details}
We first solve the proposed adversarial augmentation in (\ref{eq_positiveneighbors}) and (\ref{eq_negativeneighbors})  by the derivative-free optimization approach RACOS algorithm \cite{yu2016derivative}. The distance measure $d(\cdot)$ in (\ref{eq_positiveneighbors}) and (\ref{eq_negativeneighbors}) is set to be the angular cosine distance because of its superior discriminative information \cite{nair2010rectified}. Let $\rho = \min_{\boldsymbol{x}_i,\boldsymbol{x}_j \in \mathcal{X}_c}d(\boldsymbol{x}_i, \boldsymbol{x}_j)$, then we set the radius parameters $r_1, r_2 = \rho$, and $r_3=3\times \rho$ for $\mathcal{X}_c$. Further $T=200$ and $\gamma$ is $10^{-2}$.

Secondly, the signed graph is built upon augmented data $\mathcal{X}$. For each graph node,  $n^+$ and $n^-$ in (\ref{eq_adjacency}) are optimally chosen as 1 and 4 respectively using grid search. In addition, the metric $dist(\cdot)$ in (\ref{eqs:E_graphMatrix}) is also the angular cosine distance and $m$ is 1. 

Finally, the deep neural network is built with stacks of $3\times3$ kernel-sized separable convolutional layers. The first three blocks are equipped with increasing feature maps (128, 256, 728) and decreasing spatial squared size ($224$, $112$, $56$), and the consecutive seven blocks keep the same feature map with size $28$. After global averaging and three fully connected layers of 1024 neurons, a softmax layer is padded for label prediction. Dropout layers with $50\%$ dropout rate and weight decay with $l_2$ norm rate $10^{-4}$ are used to prevent over-fitting. Residual skips are added in order to solve the gradient diverging and vanishing problems. The regularization parameter $\lambda$ in (\ref{eqs:loss}) is optimally chosen as $1$. 

\subsection{Results and analysis}
\begin{figure}[t]
%\vskip -0.05in
\begin{center}
    \subfigure[Benign Masses]{\includegraphics[width=0.25\linewidth]{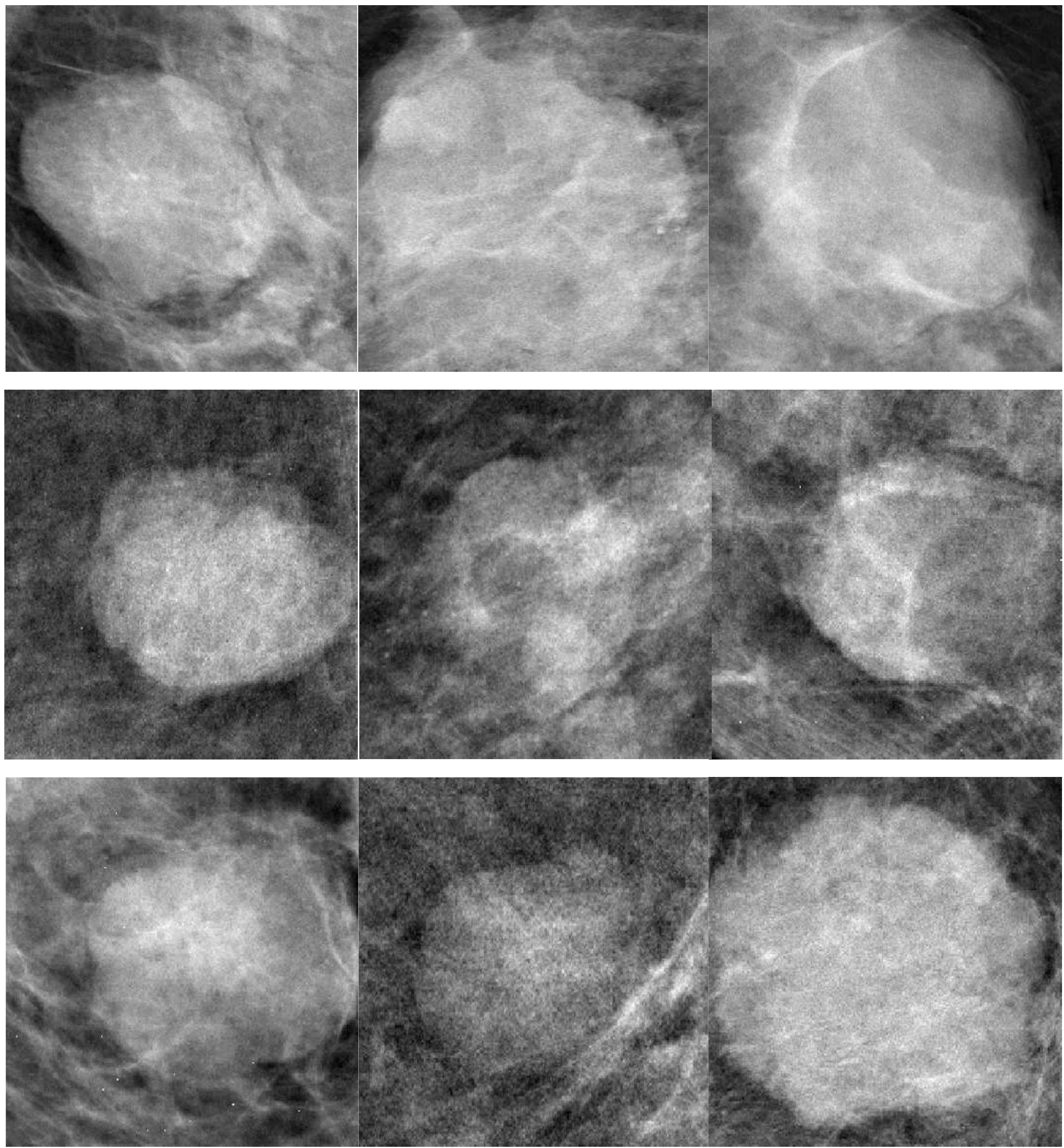}%0.48 0.82 0.06
    \label{fig:gens_b} }
    \hspace{9ex}
    \subfigure[Malignant Masses]{\includegraphics[width=0.25\linewidth]{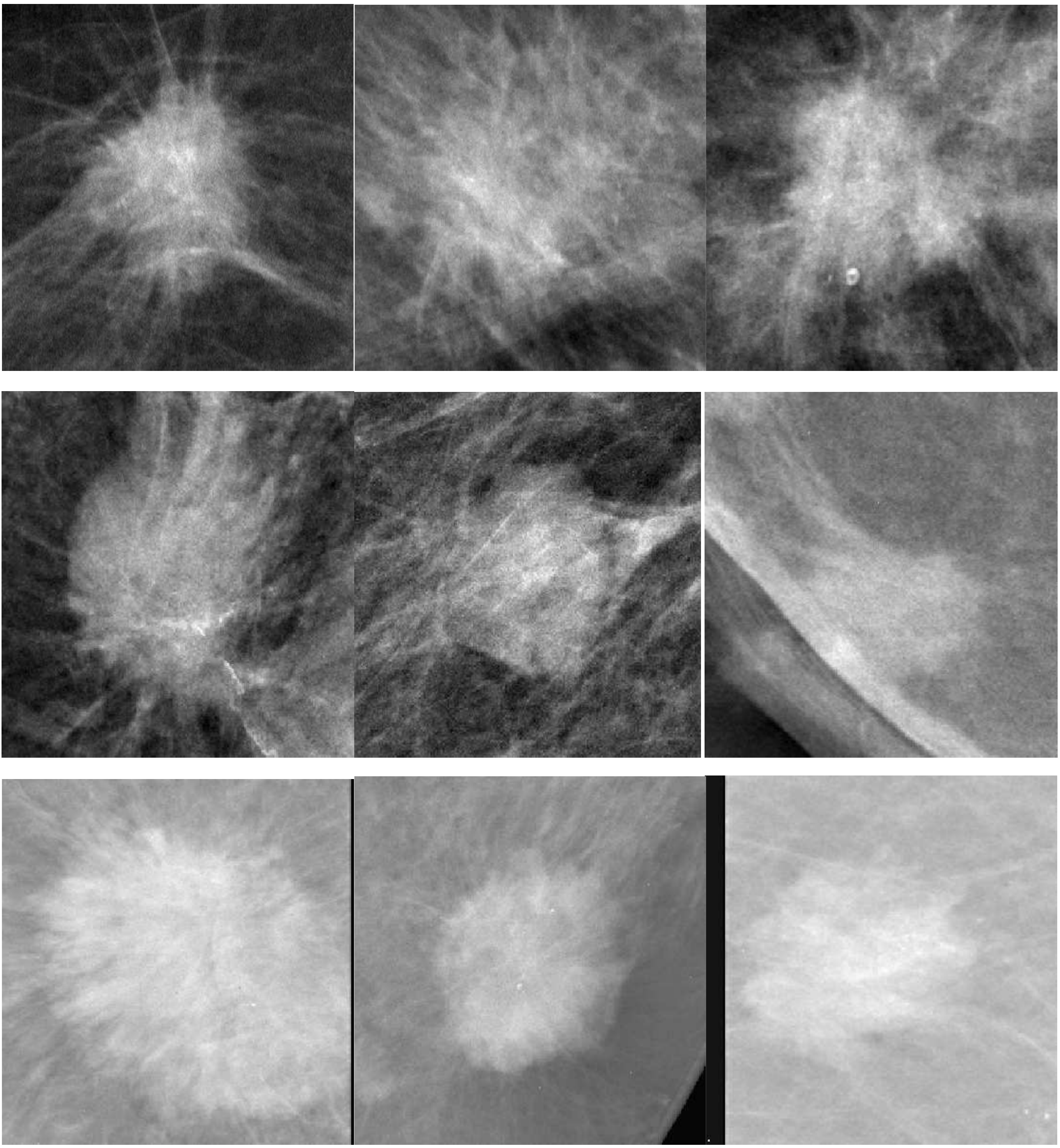}%0.35
    \label{fig:gens_m} }
\end{center}
\vskip -0.2in
\caption{Generated mammogram examples by proposed adversarial augmentation strategy. The masses in the first row of both (a) and (b) are original data, the second and the third row are generated positive and negative neighbors, respectively.}
\label{fig:gens}
\vskip -0.1in
\end{figure}
\textbf{Adversarial Augmentation:} To examine the quality of generated images by the proposed adversarial augmentation strategy, we carry out the experiment on the INbreast dataset. Fig.\ref{fig:gens} visually shows the augmented examples. It can be seen that, for either mass type, the generated positive and negative neighbors are both similar to the original data, but the negative neighbors are more different. 

\textbf{Compare to the state-of-art:} We validate \textsc{DiagNet}'s performance with accuracy and AUC (area under the ROC curve) scores. Table.\ref{table_results} compares the state-of-art algorithms, in which \cite{li2019deep} is re-implemented and the results of the remaining ones are taken from the original papers. It shows that, the \textsc{DiagNet} has achieved the state-of-art with mean accuracy 93.4\% and AUC score 0.95. When compared with the second best algorithm  \cite{shams2018deep}, the \textsc{DiagNet}'s AUC score is significantly higher with experiments on the whole dataset without any pre-processing, post-processing or transfer learning. In addition, empirical observations show that our model is robust to noise and geometric transforms, and these results are omitted due to the space limitation. 

\begin{table}[t]
%\vskip -0.1in
\renewcommand{\arraystretch}{1}
\caption{Breast Mass Diagnosis performance comparisons of the proposed \textsc{DiagNet} and relative state-of-the art methods on INbreast test set.}
\label{table_results}
\centering
\begin{tabular}{l|c|c|c}
\hline
\bfseries Methodology & \mdseries  End-to-end &\mdseries Accuracy& \mdseries AUC  \\ 
\hline
(2012) Domingues \textit{et. al} \cite{domingues2012inbreast}& \ding{53} & 89\% & N/A   \\
(2016) Dhungel \textit{et. al} \cite{dhungel2016automated}   & \ding{51} & 91\% & 0.76  \\
(2017) Zhu \textit{et. al} \cite{zhu2017deep}                & \ding{51} & 90\% & 0.89  \\
(2018) Shams \textit{et. al} \cite{shams2018deep}                                                      & \ding{51} & 93\% & 0.92 \\
 (2019) Li \textit{et. al} \cite{li2019deep}   & \ding{51} & 88\% & 0.92   \\
\hline 
proposed \textsc{DiagNet}  & \ding{51} & $\boldsymbol{93.4}\pm1.9\%$ & $\boldsymbol{0.950}\pm0.02$ \\ 
\hline 
\end{tabular}
\end{table}   

\begin{figure}[t]
\begin{center}
    \subfigure[Configurations of $(n^+, n^-)$]{\includegraphics[width=0.39\linewidth]{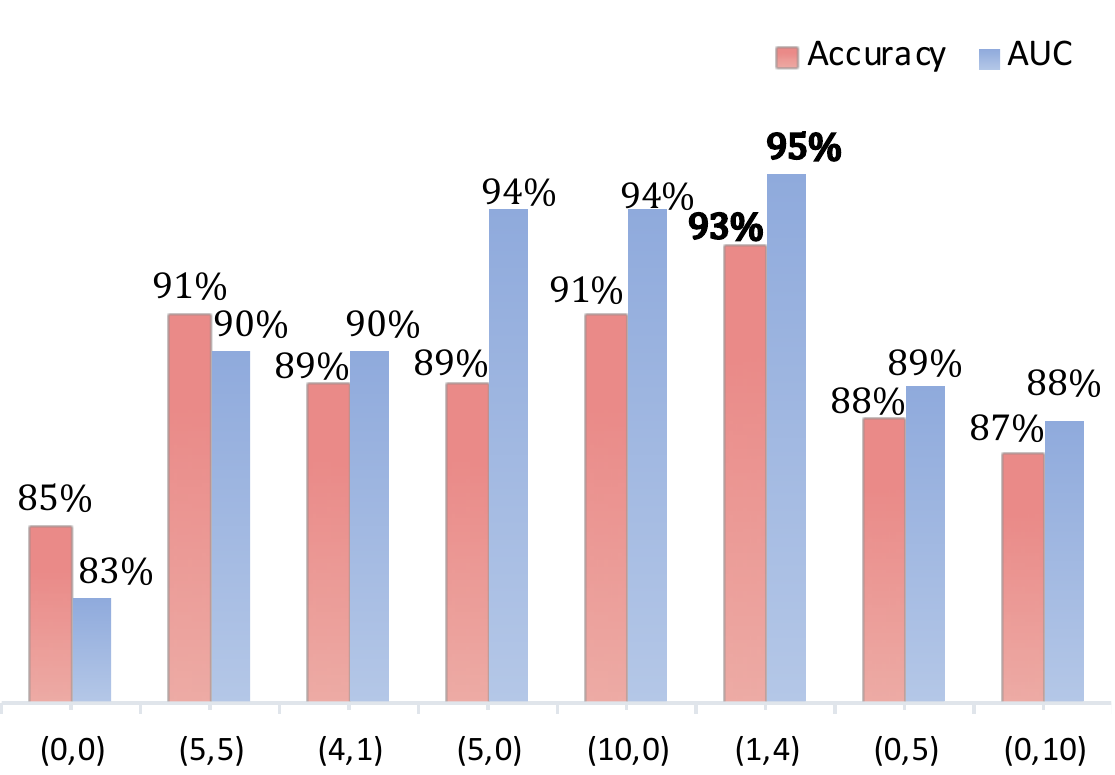}%0.48 0.82 0.06
    \label{fig:compare_n} }
    \hspace{2ex}
    \subfigure[Configurations of $\lambda$]{\includegraphics[width=0.39\linewidth]{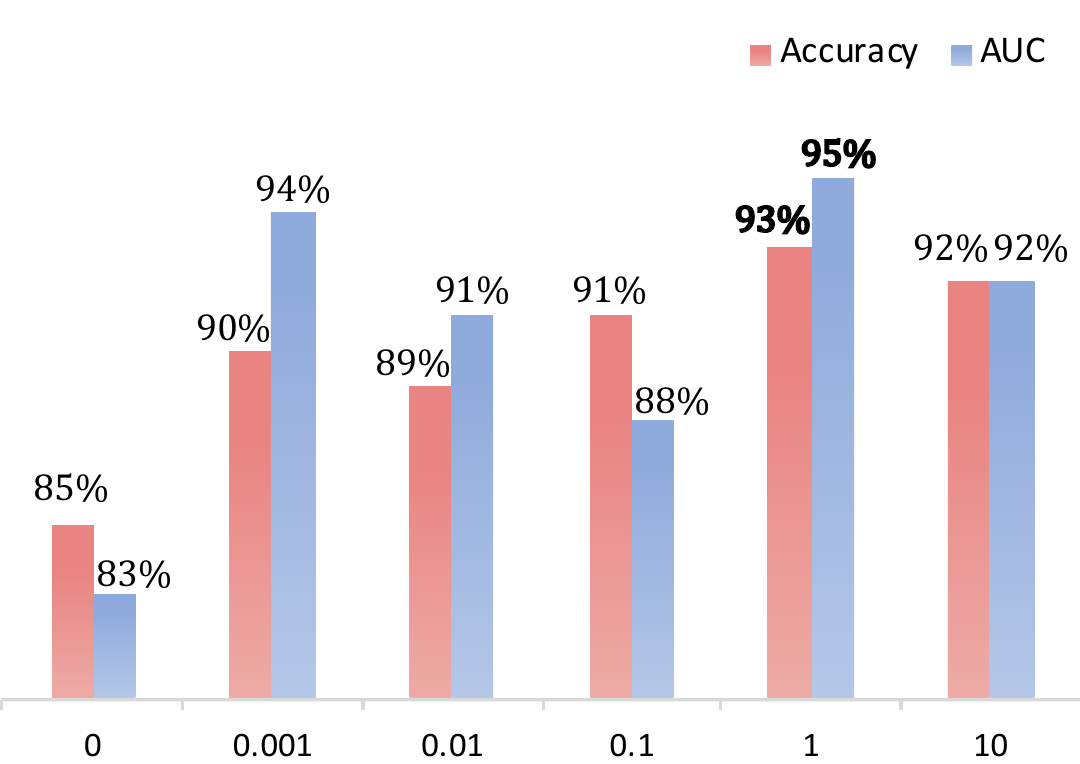}%0.35
    \label{fig:1b} }
\end{center}
\vskip -0.22in
\caption{Performance of \textsc{DiagNet} on INBreast with varying parameters. Classification accuracy and AUC score versus (a) different $n^+$ positive neighbors and $n^-$ negative neighbors and (b) various regularizer parameter $\lambda$.}
\label{fig:compare_lambda}
\end{figure}

\textbf{Importance of Signed Graph Laplacian regularizer:} Determining the optimal values of hyper-parameter is a big challenge in deep learning. To explore \textsc{DiagNet}'s performance with different signed graph configurations, the values of $n^+$ and $n^-$ are first grid searched with fixed regularization parameter $\lambda=1$, as shown in Fig.\ref{fig:compare_n}. The best performance occurs when $n^+=1$ and $n^-=4$, which increases at least by 8\% the accuracy rate and by 12\% the AUC score compared to the baseline (no graph regularization, $n^+,n^-=0$). This confirms the effectiveness of using the signed graph regularization. In addition, results show that the \textsc{DiagNet} achieves good performance only when both $n^+$ and $n^-$ are considered in the corresponding singed graph construction.  Fig.\ref{fig:compare_lambda} shows the performances with various values of $\lambda$, where the best result occurs at $\lambda=1$.

\section{Conclusions}

In this paper, we proposed a \textsc{DiagNet} for improved mammogram image analysis. By integrating the signed graph regularizer and the adversarial sampling augmentation,  \textsc{DiagNet} works in a simple but effective way to learn discriminative features. Extensive experiments show that our method outperforms state-of-the-art on breast mass diagnosis in mammography. 

\bibliographystyle{splncs04}
\bibliography{reference}
\end{document}